\documentclass[aps,twocolumn,nofootinbib,superscriptaddress,amsfonts,floatfix
]{revtex4-1} 
\usepackage{graphicx,amsmath,amssymb,amstext}
\usepackage{amssymb,amsbsy,amsfonts,amsthm,color}

\usepackage{mathtools}
\usepackage{epsfig}
\usepackage{graphicx}
\usepackage{subfigure}
\usepackage[dvipsnames]{xcolor}
\definecolor{linkcolor}{rgb}{0.0,0.3,0.5}
\usepackage[unicode, colorlinks=true, linkcolor=linkcolor, citecolor=linkcolor, 
filecolor=linkcolor,urlcolor=linkcolor, pdfusetitle]{hyperref}
\usepackage{cancel}
\usepackage{balance}
\usepackage[capitalise]{cleveref}
\usepackage{xspace}
\usepackage{enumerate}
\usepackage{ulem}
\usepackage{mdframed}
\usepackage{booktabs}
\AtBeginDocument{%
  \heavyrulewidth=.08em
  \lightrulewidth=.05em
  \cmidrulewidth=.03em
  \belowrulesep=.65ex
  \belowbottomsep=0pt
  \aboverulesep=.4ex
  \abovetopsep=0pt
  \cmidrulesep=\doublerulesep
  \cmidrulekern=.5em
  \defaultaddspace=.5em
}

\graphicspath{{Figures/}}

\usepackage{color}

\newcommand{\Beq}{\begin{eqnarray}}
\newcommand{\Eeq}{\end{eqnarray}}

\def\lsim{\mathrel {\vcenter {\baselineskip 0pt \kern 0pt \hbox{$<$} \kern 0pt \hbox{$\sim$} }}}

\newcommand{\mpl}{M_{\mbox{\tiny Pl}}}

\def\gsim{\mathrel {\vcenter {\baselineskip 0pt \kern 0pt \hbox{$>$} \kern 0pt \hbox{$\sim$} }}}

\newcommand{\grchombo}{\textsc{GRChombo}\xspace}
\newcommand{\RomanNumeralCaps}[1]

\newcommand*{\defeq}{\mathrel{\vcenter{\baselineskip0.5ex \lineskiplimit0pt
                     \hbox{\scriptsize.}\hbox{\scriptsize.}}}%
                     =}

\definecolor{mypurple}{RGB}{143, 116, 210}

\newcommand{\dd}{\mathrm{d}}

\newcommand{\bs}{\boldsymbol} 

\newcommand{\cambridge}{Centre for Theoretical Cosmology, Department
of Applied Mathematics and Theoretical Physics, University of Cambridge,
Wilberforce Road, Cambridge CB3 0WA, United Kingdom}
\newcommand{\jhu}{Department of Physics and Astronomy,
Johns Hopkins University, Baltimore, MD 21218, USA}
\newcommand{\kcl}{Theoretical Particle Physics and Cosmology
Group, Physics Department, Kings College London, Strand, London WC2R 2LS, United
Kingdom}
\newcommand{\oxford}{Astrophysics, University of Oxford, DWB, Keble
Road, Oxford OX1 3RH, United Kingdom}
\newcommand{\qmul}{School of Mathematical Sciences, Queen Mary University of London, Mile End Road, London E1 4NS, United Kingdom}

\begin{document}
{\hfill KCL-TH-2021-82}
\title{The Gravitational Afterglow of Boson Stars}

\author{Robin Croft}
\email{rc634@cam.ac.uk}
\affiliation{\cambridge}

\author{Thomas Helfer}
\email{thomashelfer@live.de}
\affiliation{\jhu}

\author{Bo-Xuan Ge}
\email{bo-xuan.ge@kcl.ac.uk}
\affiliation{\kcl}

\author{Miren Radia}
\email{M.R.Radia@damtp.cam.ac.uk}
\affiliation{\cambridge}

\author{Tamara Evstafyeva}
\email{te307@cam.ac.uk}
\affiliation{\cambridge}

\author{Eugene A. Lim}
\email{eugene.a.lim@gmail.com}
\affiliation{\kcl}

\author{Ulrich Sperhake}
\affiliation{\cambridge}

\author{Katy Clough}
\affiliation{\oxford}
\affiliation{\qmul}

\begin{abstract}
    In this work we study the long-lived post-merger gravitational wave signature of a boson-star binary coalescence.  We use full numerical relativity to simulate the post-merger and track the  gravitational afterglow
    over an extended period of time. We implement recent innovations for the binary initial data, which significantly reduce spurious initial excitations of the scalar field profiles, as well as a
    measure for the angular momentum that allows us to track the total momentum of the spatial volume, including the curvature contribution.
    Crucially, we find the afterglow to
    last much longer than the spin-down timescale.
    This prolonged \emph{gravitational wave afterglow} provides a characteristic signal that
    may distinguish it from other astrophysical sources.
\end{abstract}
\maketitle


\section{Introduction} \label{sect:intro2}

\begin{figure*}[t]
\begin{center}
\href{https://youtu.be/JE5FRG7kgvU}{
{\includegraphics[width=2.0\columnwidth]{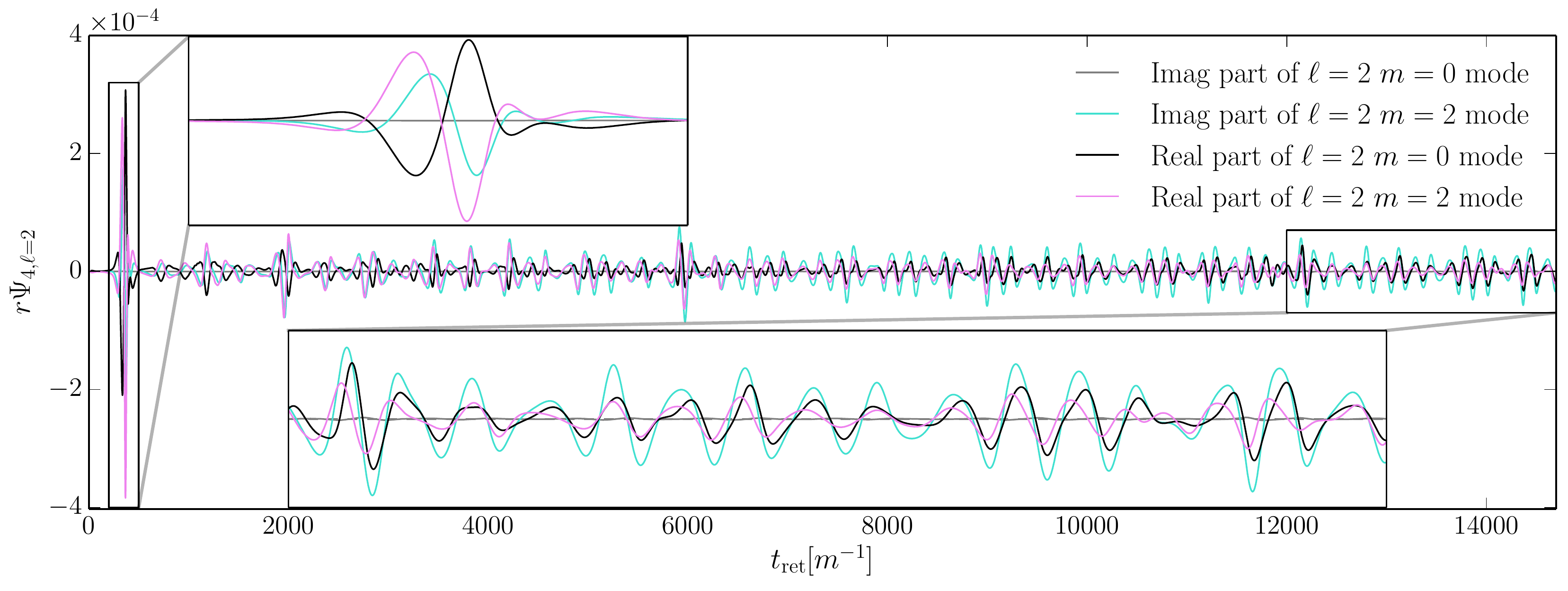}} }
\caption{{\bf Gravitational wave afterglow} emission from the collision of two boson stars with non-zero impact parameter ( ``medium'' run in \cref{tab:Overview Runs}). The Weyl scalar $\Psi_4$ is extracted at $r=220~m^{-1}$ and we only show the dominant $(\ell,m)=(2,2),(2,0)$ modes. The $(\ell,m)=(2,\pm1)$ vanishes identically due to symmetry. There is a large initial burst at merger (first zoom-in box), followed by a long, but irregular signal produced by the excited remnant boson star.  A video of the merger can be found at \url{https://youtu.be/JE5FRG7kgvU}.
 }

\label{fig:GW_signal}
\end{center}
\end{figure*}

Gravitational waves (GWs) were first predicted by Einstein in 1916 as
a consequence of general relativity.  Their recent detection by the
LIGO and Virgo observatories has opened up a new window on the
Universe. This window has led to new probes of the nature of black holes
(BHs) and to a wealth of astrophysical findings, challenging our
understanding of stellar evolution and binary population models \cite{LIGOScientific:2016aoc,LIGOScientific:2018jsj,LIGOScientific:2020kqk,LIGOScientific:2021psn}.
However, one of the most exciting and as yet unrealized prospects is
to use GWs to shed light on the nature of the dark matter (DM)
component of the Universe's energy budget. In the absence of direct
couplings between dark and baryonic matter, gravitational interactions will be the only way to
probe fundamental characteristics of DM - i.e.~mass, spin and strength
of self-interactions.  With weakly interacting massive particles (WIMPs)
proving elusive in direct detection experiments, there has been a resurgence in the interest of other DM
candidates, particularly those with low masses ($m \leq $ eV) and bosonic in nature.
Promising alternatives of this type include the QCD axion, axion-like particles
(ALPs) motivated by string theory compactifications, and ``dark
photons'' \cite{Arkani-Hamed:2008hhe,Feng:2010gw,Marsh:2015xka,Svrcek:2006yi,Kim:1986ax,Arvanitaki:2009fg,Ringwald:2012cu,Wilczek:1977pj,Peccei:1977hh,Weinberg:1977ma}.

These bosonic distributions may condense, for example from localised overdensities \cite{Widdicombe:2018oeo}, into gravitationally bound compact objects, which are referred to as boson stars (BS) \cite{Amin:2018xfe,PhysRev.172.1331,PhysRev.187.1767,PhysRev.97.511,Visinelli:2021uve,Liebling:2012fv,Schunck:2003kk,Guerra:2019srj,Vaglio:2022flq,Cardoso:2022vpj,Amin:2021tnq,Zhang:2020bec}. Stationary equilibrium solutions of this type have been
found for different types of bosons, including scalars \cite{Sanchis-Gual:2020mzb,Alcubierre:2021psa,Boskovic:2021nfs,Astefanesei:2003qy,Kaup:1968zz,Ruffini:1969qy,Colpi:1986ye,Lee:1986ts,Friedberg:1986tq,Schunck:1999zu,Jetzer:1989av,Pugliese:2013gsa,Hawley:2002zn,Urena-Lopez:2012udq,Muia:2019coe,Helfer:2016ljl,Alcubierre:2018ahf,Coleman:1985ki,Guerra:2019srj,Kleihaus:2005me,Kleihaus:2011sx,Cardoso:2014sna,Amin:2014fua,Amin:2011hj,Amin:2010dc}, vector fields \cite{Brito:2015pxa,Minamitsuji:2018kof,Brito:2015yga,Zhang:2021xxa,CalderonBustillo:2022cja,March-Russell:2022zll,Gorghetto:2022sue,Herdeiro:2021lwl,Bustillo:2020syj,Minamitsuji:2017pdr,Sanchis-Gual:2017bhw,Duarte:2016lig,SalazarLandea:2016bys,Zilhao:2015tya} or higher-spin fields \cite{Jain:2021pnk}.

Binary coalescences involving BSs
represent a promising channel to observationally
identify or constrain their populations. Their potentially high compactness implies that mergers can generate GWs detectable with present GW observatories. Most present work in the literature on BSs focusses on the GW signatures generated during the pre-merger infall or inspiral
\cite{Herdeiro:2020kba,Cardoso:2017cfl,Sennett:2017etc,Diamond:2021dth,Pacilio:2020jza} and during the
merger phase itself \cite{Amin:2020vja,Sanchis-Gual:2018oui,Sanchis-Gual:2020mzb,Sanchis_Gual_2019,Widdicombe:2019woy,
Helfer:2018vtq,Palenzuela:2017kcg,Bezares:2018qwa,Liebling:2012fv,Choptuik:2009ww,Palenzuela:2007dm,Bezares:2017mzk,
Palenzuela:2006wp,Dietrich:2018jov,Dietrich:2018bvi,Jaramillo:2022zwg,Bezares:2022obu,Macedo:2013jja}; these are, of course, the regimes of most notable
interest in the GW observation of neutron-star and black-hole
binary coalescences. The main focus of our work, however, is
the long-lived post-merger GW emission or {\it afterglow} resulting
from the merger of two BSs into a single compact but horizon-free
remnant (see \cref{fig:GW_signal}).
First indications of such an afterglow were noted in Ref.~\cite{Helfer:2018vtq} in the case of a head-on collision resulting in a highly perturbed BS. Here, we demonstrate that this afterglow can be very long lived, with barely any decay in amplitude following a transient burst during the merger phase itself. The characteristics of this post-merger afterglow contrast sharply with the corresponding GW signatures of most BH or NS mergers, which, if
resulting in BH formation, are dominated by the exponential quasi-normal ringdown. 

We illustrate and explore in detail the gravitational afterglow
of BSs for the case of the inspiral and merger 
of two equal-mass BSs in a collision with a non-zero impact parameter. For the moderate compactness of the initial binary constituents chosen in our simulations, the final state of the collision
is a highly perturbed BS with decreasing spin. Crucially, this spin-down occurs on a time scale {\it much longer} than a single GW oscillation time period.  The associated long-lived GW afterglow may exhibit information about the post-merger dynamics of such systems. In particular, we find an intriguing correlation between the phases of different GW multipoles and the dynamical spin amplitude. 

The results of this paper suggest that using standard merger templates consisting mostly of the inspiral and merger contributions may be insufficient to capture fundamental dynamics of a boson-star merger event. Rather, comprehensive BS searches likely require extended waveform templates which also capture the rich post-merger GW afterglow phenomenology.

This paper is structured as follows: In \cref{section:inital_data} we 
briefly summarize our computational framework, list the parameters
of our initial configurations and the grid setups employed in their
time evolution, and introduce the diagnostics specific
to our simulations. In Sec.~\ref{sec:Remannt},
we list the key features of the post-merger
remnant. The corresponding GW signal is discussed in more detail
in Sec.~\ref{sec:GWs} and we conclude in Sec.~\ref{sec:conclusion}.
Technical details of the numerical methodology, the calculation
of angular momentum, and the estimate of numerical uncertainties
are relegated to Appendices \ref{appendix:numerical_methodology},
\ref{sec:methods} and \ref{sec:numacc}.
Unless explicitly stated otherwise,
we use units where the speed of light and
Planck's constant are set to unity, $c=\hbar=1$, and we express the gravitational constant in terms of the Planck mass $G = 1/\mpl^2$. Unless specified otherwise, Latin indices run from 1 to 3 while Greek ones run from 0 to 3.

\begin{figure*}[t]
\begin{center}
\href{https://youtu.be/JE5FRG7kgvU}{
    {\includegraphics[width=2\columnwidth]{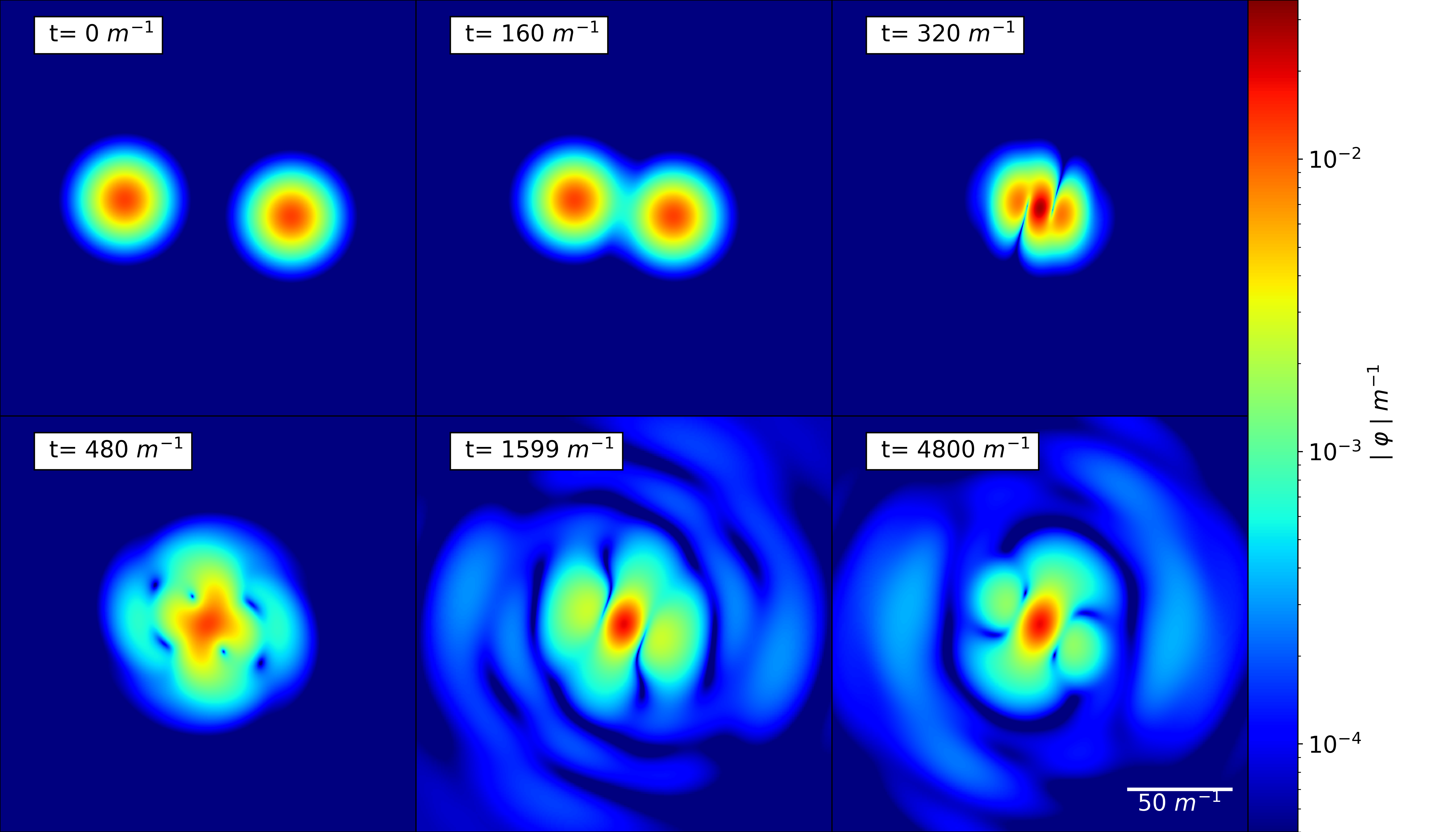}}
    }
\caption{
    Snapshots of the scalar field amplitude $|\varphi|$
    in the orbital plane for a
    grazing collision of two boson stars
    of equal mass $M=0.395~\mpl^2 m^{-1}$ starting with initial horizontal distance $80~m^{-1}$, impact
    parameter $b=8~m^{-1}$ (vertical center to center distance) and initial velocity $v=\pm 0.1$
    in the $x$ (horizontal) direction. A video of the merger can be found at \url{{https://youtu.be/JE5FRG7kgvU}}}.
\label{fig:Panellog}
\end{center}
\end{figure*}

\section{Simulation Set-up}\label{section:inital_data} 

Throughout this work, we model BSs as a complex scalar field
minimally coupled to the gravitational sector of a Lorentzian
manifold with metric $g_{\alpha\beta}$. The corresponding
Lagrangian is given by the Einstein-Hilbert action plus a
matter term,
\begin{eqnarray}
  &&\mathcal{L} = \frac{1}{16\pi G}R + \mathcal{L}_{\rm m},
  \\[10pt]
  &&\mathcal{L}_{\rm m} =-\frac{1}{2}g^{\mu\nu}\nabla_{\mu}{\bar{\varphi}}\nabla_{\nu}{\varphi}-\frac{1}{2} V(\varphi),
\end{eqnarray}

with the potential function for a non-interacting scalar
of mass $m$,
\begin{equation}
    V(\varphi) = m^2 \bar{\varphi} \varphi.
\end{equation}
This choice of potential results in BS solutions that are referred to as {\it mini-boson stars}
\cite{ Kaup:1968zz,Gleiser:1988rq,Jetzer:1988vr}. 
Our construction of boson-star binary initial data can loosely
be summarized in the following three steps.
\begin{enumerate}[(i)]
    \item Generate a stationary, non-rotating solution for a single boson star.
    \item Apply a Lorentz boost to obtain a single star with linear momentum.
    \item Superpose two such solutions
    according to the procedure described in
    Refs.~\cite{Helfer:2018vtq,Helfer:2021brt} which
    substantially reduces spurious initial oscillations of the
    individual BSs as compared to the more common procedure
    of plain superposition.
\end{enumerate}

Most of our results are obtained from
simulating a grazing collision of two stable BSs, each with mass\footnote{This mass is obtained for a central scalar-field amplitude $|\varphi(0)|/\mpl =0.0124$ and results in a compactness estimate
$\mathcal{C} \defeq \max (m(r)/r) = 0.024$ in radial gauge.
For comparison the Kaup limit configuration has
$M=0.633$ and $\mathcal{C} = 0.12$.}  $M = 0.3950
~\mpl^2 m^{-1}$ and
initial velocity $v = \mp(0.1,0,0)$.

The stars are initially located $d_{\text{init}} = 80~m^{-1}$
apart in the $x$ direction and also offset by an impact parameter
$b=8~m^{-1}$ perpendicular to this axis; it is through this offset
(rather than a velocity component off the $x$ direction) that
the binary is endowed with initial orbital angular momentum.
The Newtonian point-particle estimate for the angular momentum
of this configuration,
\begin{equation}
    L_{ \rm N} =  M b v_x = 0.316~\mpl^2/m^2,
\end{equation}
agrees remarkably well with the relativistic measurement which only deviates by 1.1 \%.
A summary of this binary's initial data together with the main parameters of the numerical setup are given in Table \ref{tab:Overview Runs}. We have simulated numerous other
binary configurations -- different boson star masses, initial velocities $v$ and impact parameters $b$ -- that display qualitatively
the same behaviour. The main features of the binary dynamics
that we will report in the following are thus {\it not} a
consequence of any fine tuning of initial data.

For all simulations, we use a square box of width $D = 1024~m^{-1}$, employing the adaptive mesh refinement (AMR) capabilities of GRChombo \cite{Radia:2021smk,Andrade:2021rbd,Clough:2015sqa}. Besides the standard computation of the Newman-Penrose scalar
whose implementation in GRChombo is described in detail in
Ref.~\cite{Radia:2021smk}, we compute in our simulations
two diagnostic quantities specific to the BS systems under study.

First, we introduce the mass measure 
 \begin{equation}
    M = \int_{\Omega} \rho \sqrt{\gamma} dV~,\label{eqn:Massmeasure}
\end{equation}
where $\rho = T_{\mu\nu}n^{\mu}n^{\nu}$ is the energy density as measured
by observers moving along the normal vector $n^\mu$ to the
spatial hypersurfaces. 
The second
is a time dependent
measure $\tilde{L}$, defined in Eq.~(\ref{eqn:DefAngMom}),
for the angular momentum contained
inside a specified volume $V$. This quantity is obtained
by adding to the initial angular momentum the time integrated
rate of change due to the source of momentum that crucially
includes contributions from the spacetime dynamics; the details
for computing this quantity $\tilde{L}$ are given in
Appendix \ref{sec:methods}.

\section{The Merger Remnant}\label{sec:Remannt}

\begin{figure*}[t]
\begin{center}
{\includegraphics[width=2.0\columnwidth]{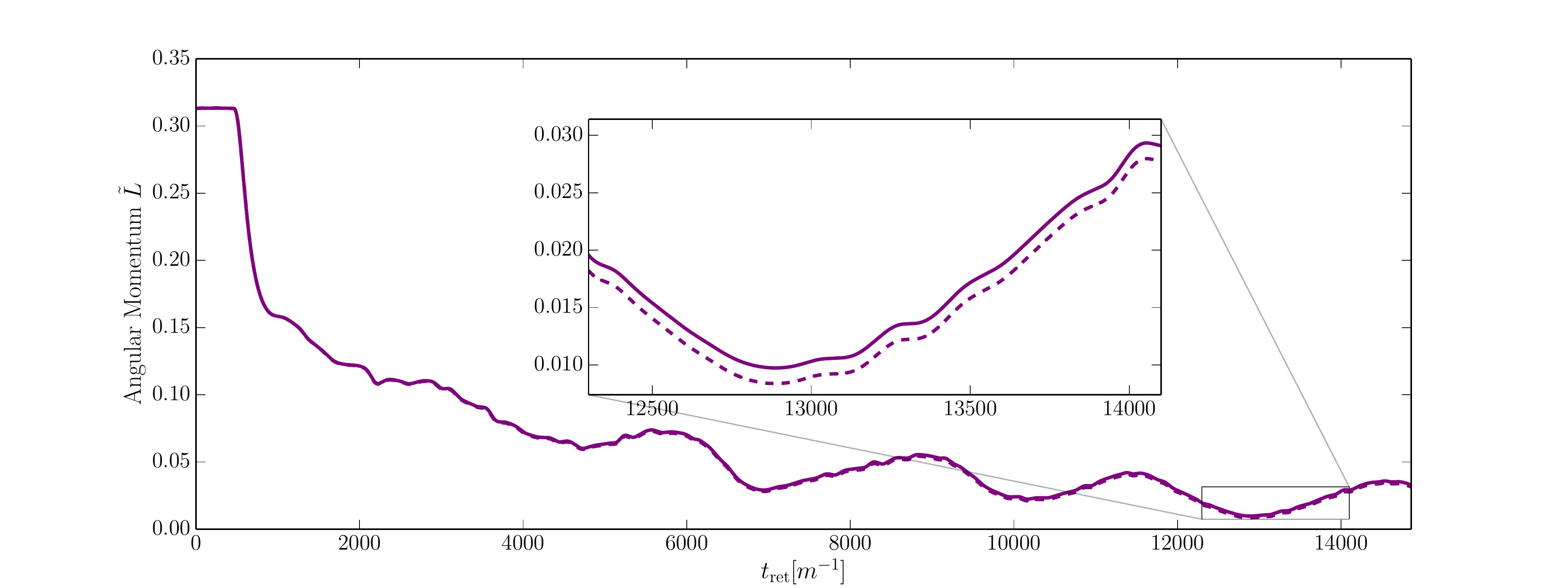}}
\caption{
    {\bf Angular momentum of the scalar field}: We show the angular
    momentum $\tilde{L}$ (see definition in  \cref{eqn:DefAngMom}) inside a coordinate sphere of radius $60~m^{-1}$ as a function of time. We compute
    $\tilde{L}$ in two ways, (i) via integrating the outgoing flux -- solid lines (see \cref{eqn:DefAngMomFlux}) -- and (ii) as a volume-integral --
    dotted lines. This was plotted using run 2 of \cref{tab:Overview
    Runs}}
\label{fig:AngMom}
\end{center}
\end{figure*}
When colliding two BSs with angular momentum, we
expect one of the following outcomes:
\begin{enumerate}
    \item A toroidal spinning BS
        \cite{PhysRevD.90.024068,Yoshida:1997qf,1996rscc.conf..138S,Siemonsen:2020hcg},
    \item A non-spinning BS with perturbations carrying away the angular
        momentum
        \cite{Macedo:2013jja,Macedo:2016wgh,Yoshida:1994xi,Flores:2019iwp},
    \item A black hole \cite{Helfer:2018vtq,Palenzuela:2017kcg,Bezares:2018qwa}, or,
    \item Total dispersion of all matter.
\end{enumerate}
For sufficiently small compactness of the progenitors the merger does not form a black hole. While we have observed black-hole
formation in some of our calibration runs starting with more compact
BSs, in the remainder of this paper we focus on the scenario
where the merger results in a compact bosonic configuration
without a horizon as shown in Fig.~\ref{fig:Panellog}.
The scalar-amplitude profiles in this figure (nor at any other
times during the evolution) display no signs
of a toroidal structure and we therefore interpret the merger
outcome as a perturbed non-spinning BS corresponding to the
second item in the above list; cf.~also
Refs.~\cite{Palenzuela:2017kcg,Bezares:2018qwa}.

In Fig.~\ref{fig:AngMom}, we display
the angular momentum $\tilde{L}$ of the BS configuration
inside a coordinate sphere of radius $60~m^{-1}$ throughout
inspiral, merger and the afterglow phase. Up to the time
of merger around $t\approx 300~m^{-1}$, the angular momentum
remains approximately constant before rapidly decreasing
in the post-merger phase. To leading order,
the tail of the resulting curve
$\tilde{L}(t)$ is approximated by an exponential decay with
half-life $4\times10^3~m^{-1}$, as obtained from an exponential
fit to the data of Run 2 starting at $t=2\,000~m^{-1}$.

Translated into SI units, the half-life is
\begin{equation}
    t_{\text{half}} = 83 \text{ years}  \left(\frac{10^{-21}
    \text{eV}}{m}\right)~\label{eqn:decay_time}.
\end{equation}

For a scalar mass $m =  10^{-14}$ eV, for example,
the dominant frequency of the $\ell=2$, $m=0$ signal
falls into the most sensitive region of the LISA noise curve (see \cref{eqn:freqGW} below) and we obtain a half-life of $\sim4~{\rm min}$. For scalar masses in or above this regime, this implies
that a delayed formation of a black hole, should it occur,
will result in a black hole with negligible spin. With regard to
the possibility of the formation of a black-hole population
through isolated BS progenitors
\cite{Helfer:2016ljl,Muia:2019coe}, this implies that
spinning black holes are unlikely to have formed this way unless
the BS progenitors are composed of ultra light scalar particles.
More quantitatively, we see from
Eq.~(\ref{eqn:decay_time}), that
astrophysically large decay times for the angular momentum
of order
$\mathcal{O}({\rm Myr})$ require ultra light
scalars with mass\footnote{Note that candidates below $m\lesssim 10^{-22}\,{\rm eV}$ are ruled out as constituting 100\% of the DM by structure formation constraints but may still form some proportion of the DM \cite{Marsh:2015xka}}
$m\lesssim 10^{-25}\,{\rm eV}$ .

The rapid drop in the angular momentum of rotating
scalar soliton stars has been noticed as early as the mid 1980s
\cite{Lee:1986ts,Friedberg:1986tq},
but we note that the post-merger evolution of our $\tilde{L}$,
besides an approximately exponential drop, also exhibits
significant oscillations on a time scale of about
$2\,000~m^{-1}$. We conjecture that these oscillations
arise from the complex dynamics of the post-merger remnant
and may carry memory of its formation process.

\begin{figure}[t]
\begin{center}
    {\includegraphics[width=1.0\columnwidth]{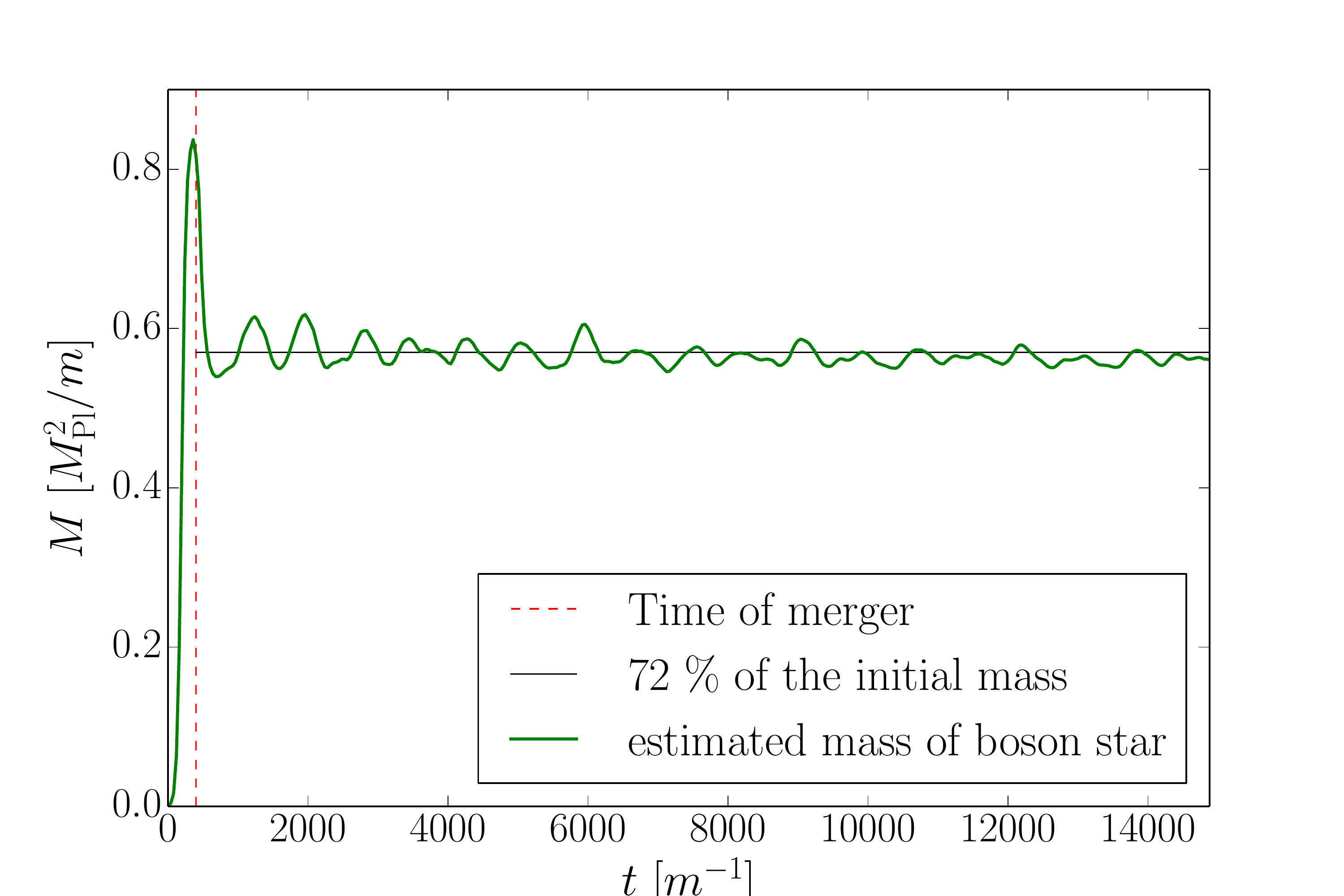}}
    \caption{{\bf The estimated mass} using \cref{{eqn:Massmeasure}}
    contained inside a box $\Omega$ with side length $40~m^{-1}$ as computed for the ``medium'' 
    resolution run 2 of Table \ref{tab:Overview Runs}. 
    Since the BSs are
    not initially inside this box, the mass at $t = 0$ is close to zero. The small fluctuations after merger
   are due to the gauge-dependence of the measure.
   }
\label{fig:energy}
\end{center}
\end{figure}

We also observe significant oscillations in the time evolution
of the merger remnant's mass $M$ as defined in Eq.~(\ref{eqn:Massmeasure}). As demonstrated in
Fig.~\ref{fig:energy}, however, the mass evolution differs
significantly from that of the angular momentum.
First, the mass gradually levels off
at $M \sim 0.57~\mpl^2/m$ or $\sim 72$ \% of the initial mass instead of decaying over time.
Second, the oscillations occur on a much shorter time scale.

\section{Gravitational wave signal}\label{sec:GWs}

We now turn our attention to the GW signal generated by the
BS coalescence. We find this signal to be dominated by the
$(l,m)=(2,\pm 2)$ and $(2,0)$ quadrupole modes which are
displayed in Fig.~\ref{fig:GW_signal} for Run 2 using an extraction
radius $220~m^{-1}$. The large burst around merger at $t\approx300~m^{-1}$ (see the upper left inset of the figure)
closely resembles the corresponding features
regularly seen in the merger of black-hole binaries.
The ensuing long-lived, semi-regular radiation clearly visible
with barely any signs of diminuition up to the end of our
simulation, however, drastically
differs from the familiar ringdown of a merged black hole. This
{\it afterglow} signal is the main result of our study. We
emphasize that this signal is well resolved (rather than
merely displaying numerical noise), and also persists with
negligible variation under changes in the numerical resolution
of our grid. As discussed in more detail in
\ref{sec:numacc}, we estimate the numerical uncertainty
of the $r\Psi_4$ signal at about $7\,\%$ during the afterglow phase
with most of this error budget being due to the finite extraction radius.
The GW signals of the higher-resolution Runs 3 and 4, if added
to Fig.~\ref{fig:GW_signal}, would almost overlap with
that shown in the figure for Run\footnote{Run 2 is our longest simulation
and therefore used for most of our analysis.} 2;
cf.~also Fig.~\ref{fig:Convergence} below.
\begin{table}
\begin{center}
\begin{tabular}{ c  c  c  c  c  c  c  c  c}
     \toprule
      & Run & $N$      & $d_{\text{init}}~[m^{-1}]$   & $b~[m^{-1}]$  & $v_x$ &  $M~[\mpl^2~m^{-1}]$  \\ 
     \midrule
     low &  1    & 256 & 80 & 8  & 0.1 & 0.395(0) \\
     medium &  2   & 320 & 80 & 8 & 0.1 & 0.395(0) \\
     high &  3    & 384 & 80 & 8 & 0.1 &  0.395(0)\\
     ultra-high &  4    & 448 & 80 & 8 & 0.1 &  0.395(0) \\
     \bottomrule
\end{tabular}
\caption{{\bf Overview} of the simulations. Here, $M$ is the individual mass of each boson star, $v_x$ the initial velocity, $b$ the impact parameter, $d_{\rm init}$ the initial distance in the $x$ direction, $b$ the vertical offset or {\it impact parameter} and $N$ is the number of cells on the coarsest AMR Level (which sets the resolution of the respective runs). We allow for seven extra refinement levels. The data associated with these runs can be found here: \url{https://github.com/ThomasHelfer/BosonStarAfterglow},}
\label{tab:Overview Runs}
\end{center}
\end{table}

\begin{figure}[t]
\begin{center}
{\includegraphics[width=1.0\columnwidth]{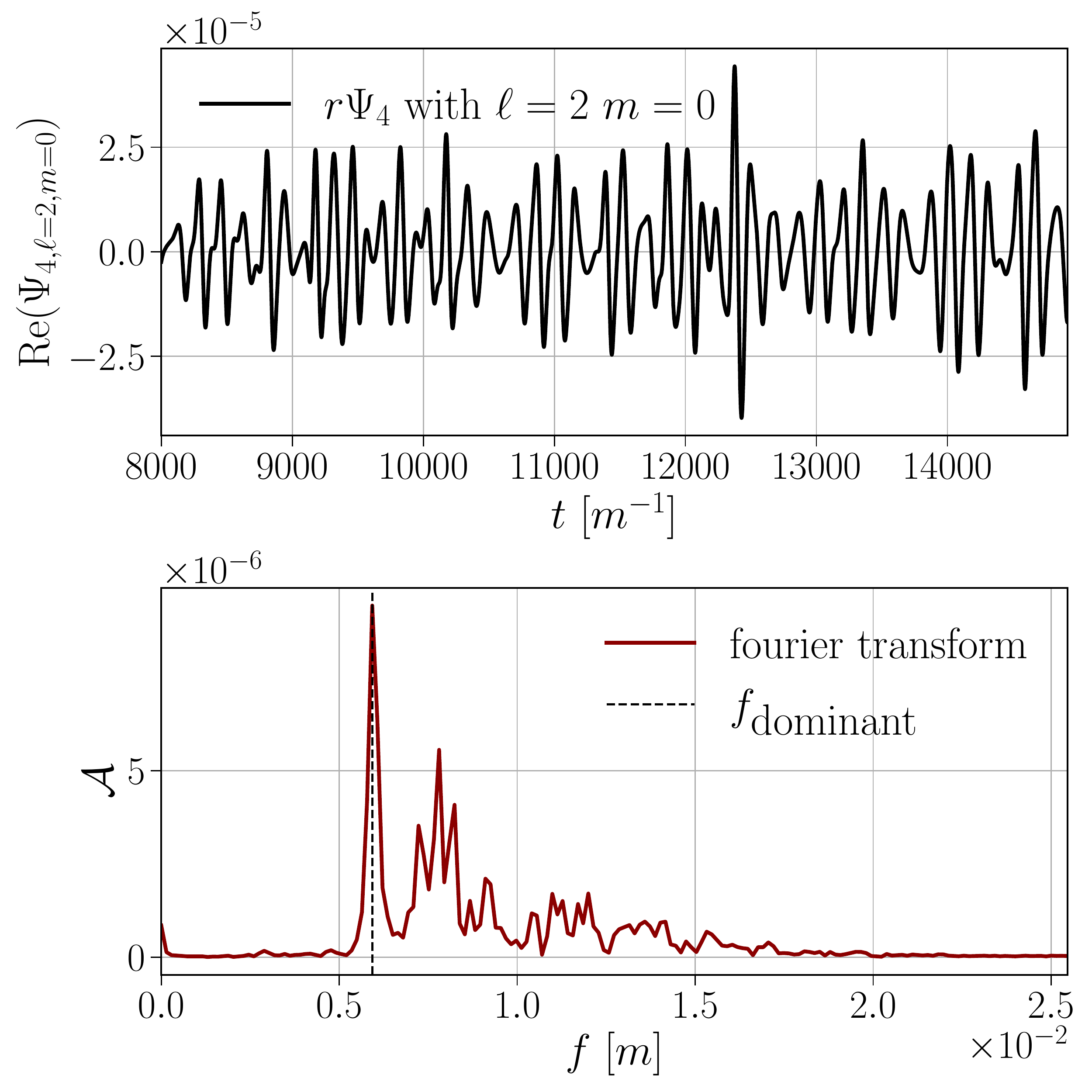}}
\caption{{\bf Time domain signal and
    Fourier transform of the $(2,0)$ mode
    of $r\Psi_4$}: We perform a Fourier
    transformation of the tail of the gravitational wave signal of the ``medium'' resolution run in \cref{tab:Overview Runs}.
    We find excellent agreement between the displayed
    spectrum for the $(2,0)$ mode and the corresponding Fourier transform of the $(2,2)$ mode; in particular, both yield the same peak frequency. 
    }
\label{fig:Fourier}
\end{center}
\end{figure}

The afterglow signal (without the prodigious merger burst)
is also shown in Fig.~\ref{fig:Fourier} together with 
its Fourier spectrum. The frequency spectrum demonstrates
contributions on many time scales, but also reveals a
narrow dominant peak at $f_{\rm dom}\approx
0.6\times 10^{-2}~m$ which, translated into SI units, can
be written as
\begin{equation}
    f_{\textrm{dom}} \sim 9.0 \cdot 10^{-2}\text{ Hz} \left( \frac{m}{10^{-14}~\text{eV}}
    \right)~. \label{eqn:freqGW}
\end{equation}
Both the time- and frequency-domain signals exhibit signature
of beating effects: the amplitude of the rapid oscillations
itself undergoes a modulation at lower frequency.

\begin{figure}[t]
\begin{center}
{\includegraphics[width=1.0\columnwidth]{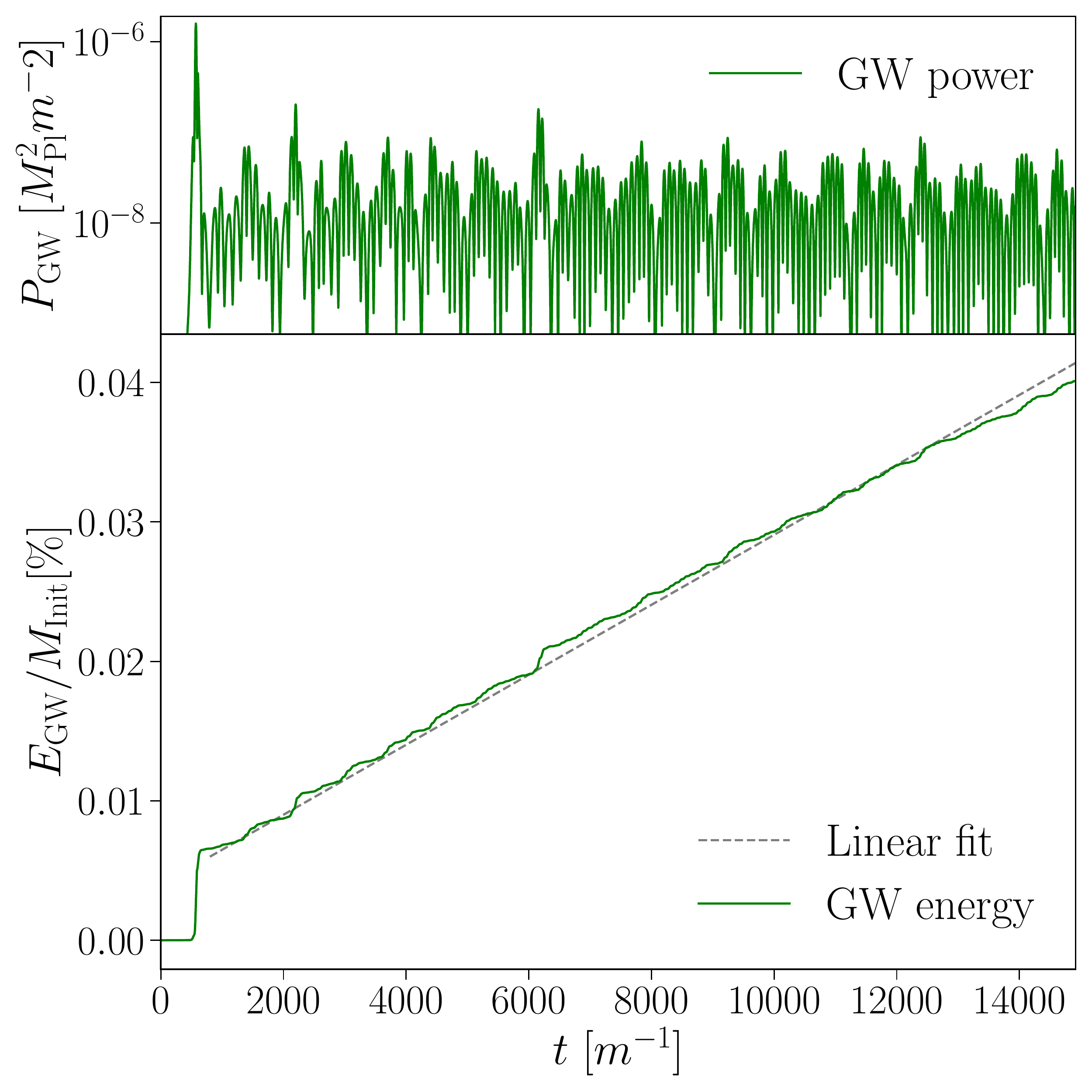}}
\caption{{\bf Radiated GW energy over time}: We calculate the energy and power radiated in gravitational waves from the ``medium'' resolution run of \cref{tab:Overview Runs}. 
We observe no significant reduction in the GW radiation over the simulation time, allowing us to estimate a lower bound on the half-life of the signal.
    }
\label{fig:EnergyPower}
\end{center}
\end{figure}

The prolonged afterglow furthermore accumulates a non-negligible
amount of energy emitted in GWs. By the end of our simulation
at $t\approx 15\,000~m^{-1}$, the radiated energy computed
according to Eq.~(\ref{eq:GWpower}) including
infall and merger has reached
$(0.04 \pm 0.0014)~\% $ of the initial mass,
corresponding to an average rate of $2.5 \times 10^{-8} M_{\rm init}~m$ (see dotted line in \cref{fig:EnergyPower}). 
The radiated energy and power are shown
as functions of time in Fig.~\ref{fig:EnergyPower}
and clearly show an approximately linear increase in
$E_{\rm GW}$ during the afterglow phase. 
This significant amount of post-merger GW emission in itself is a
striking signature of exotic binary merger progenitors
that distinguishes them from BH binaries devoid 
of significant post-merger radiation beyond the quasi-normal ringdown. 
By using windowing of the GW signal, we find that the rate of radiation in the afterglow (excluding the merger peak) decreases by about $20 \%$ over the course of the simulation.
Note that the decay in GWs is much more protracted than the drop in the angular momentum displayed in Fig.~\ref{fig:AngMom}.
Clearly, the system loses angular momentum much more rapidly
than energy.

A more subtle feature in the post-merger signal is revealed in
the multi-polar decomposition of the quadrupole signal;
more specifically in the relative position of the local extrema
in the $(\ell,m)=(2,2)$ and $(2,0)$ modes. As exhibited by
the upper left inset of Fig.~\ref{fig:GW_signal}, the amplitudes
of the $(2,2)$ and $(2,0)$ modes are almost exactly
in anti-phase around merger and remain so in the early afterglow
around $t\sim 1\,000~m^{-1}$. At late times
$t\gtrsim 3\,000~m^{-1}$, however, the
two modes are almost synchronized with their extrema
in good overlap.
The timing of this synchronization coincides remarkably well
with the drop in angular momentum shown in Fig.~\ref{fig:AngMom}
and we hypothesize the two effects are causally related.
This would imply a concrete observational signature of the
BS angular momentum in the emitted GW afterglow signal.

In physical terms, the GW afterglow is a direct consequence
of the the presence of matter around the compact merger remnant
and the resulting complex matter dynamics following the violent
merger. A qualitatively similar behaviour may arise in the
merger of neutron stars provided these do not promptly merge
into a black hole. Two key differences between neutron-star
and boson-star binaries, however, may aid considerably in the
distinction between neutron-star and BS
signals. The first consists in the
extremely long-lived nature of the BS afterglow which we
anticipate will last for much longer times than are presently
within grasp of our numerical studies; cf.~again Fig.~\ref{fig:GW_signal} and the barely perceptible
drop in the GW signal. The second fundamental discriminator
arises from the scale-free nature of the BS spacetimes;
the scalar mass parameter $m$ appears as a characteristic
scale in all dimensional variables of the GW analysis.
While NS masses are restricted to be below the Chandrasekhar limit of about $2 M_{\odot}$, BSs may theoretically exist
across the entire mass spectrum and barring for a remarkable
coincidence in the scalar mass value, will be distinguishable
from their neutron-star counter parts by the frequency regime
of their GW emission. Put the other way round, comprehensive
observational searches for GW signatures from BSs require
scanning over a wide range of frequencies using vastly
different detectors such as LIGO-Virgo-KAGRA, LISA,
third-generation detectors but also high frequency GW observatories presently under development \cite{Aggarwal:2020olq,Badurina:2019hst}.

\section{Conclusion}\label{sec:conclusion}

In this paper, we have shown that the inspiral and
coalescence of BS binaries into a non-BH remnant can produce a long-lasting GW {\it afterglow}. This signature is salient, and markedly differs in duration and -- possibly -- also
frequency from the GW signatures of more traditional astrophysical compact object mergers; it thus represents a distinct detection channel for exotic compact objects
in compact-binary-coalescence and continuous-GW searches \cite{LIGOScientific:2019yhl,KAGRA:2021una,LIGOScientific:2021jlr,LIGOScientific:2021oez,LIGOScientific:2022lsr,LIGOScientific:2022pjk,LIGOScientific:2021hvc,LIGOScientific:2021ozr}.

There are several implications resulting from our
findings. In terms of search strategies, as mentioned in the introduction, these signatures are likely to be missed if we focus exclusively on constructing pre-merger inspiral and merger waveform templates. The systematic construction of waveform templates for post-merger signatures of this type of binaries
is in its infancy at present and an immediate challenge
for further work consists in identifying an effective
parameterization of the GW signatures. Our results furthermore
demonstrate an efficient loss of angular momentum in BS mergers
resulting in a horizon-less remnant, consistent with previous studies noticing that the spin of rotating BSs decays with a fairly short half-life of $4\times 10^3~m^{-1}$ \cite{Sanchis-Gual:2019ljs}. We also observe
a remarkable correlation between the BS remnant's spin-down
in Fig.~\ref{fig:AngMom} with a gradual synchronization
of the local extrema in the GW amplitudes of the $(2,2)$
and $(2,0)$ modes; from near anti-alignment of the
peaks around merger and shortly thereafter, the extrema
gradually shift into approximate overlap over a time
interval $\Delta t \approx 2\,000~m^{-1}$ (see
Fig.~\ref{fig:GW_signal}), coinciding
exactly with the time during which the angular momentum
drops to a negligible level.
We tentatively conclude that through this synchronization,
the GW afterglow carries important information about the remnant's dynamical evolution. 

Given the extraordinary length of the afterglow signal, one
would expect the radiation from numerous BS merger events
-- if they occur -- to result in a stochastic background.
Such a background could be searched for additionally
to that expected from more traditional binary mergers
\cite{Croon:2018ftb}. Evidently, more exploration of the
underlying BS parameter space and the resulting afterglow
phenomenology will be required to relate theoretical
estimates of the GW background to hypothesized BS populations.
We reiterate, however, that nothing about our BS configurations
has been fine-tuned, so that we expect the afterglow to be a
rather generic feature of BS coalescences as long as these do 
not promptly form a black hole.

\acknowledgments We thank Caio Macedo and Emanuele Berti for helpful discussion.
T.H. is supported by NSF Grants No. AST-2006538, PHY-2207502, PHY-090003 and PHY20043, and NASA Grants No. 19-ATP19-0051, 20-LPS20- 0011 and 21-ATP21-0010. This research project was conducted using computational resources at the Maryland Advanced Research Computing Center (MARCC).
KC acknowledges funding from the European Research Council (ERC) under the European Unions Horizon 2020 research and innovation programme (grant agreement No 693024), and an STFC Ernest Rutherford Fellowship project reference ST/V003240/1.
This work has been supported by
STFC Research Grant No. ST/V005669/1
``Probing Fundamental Physics with Gravitational-Wave Observations''.
RC and TE are supported by the Centre for Doctoral Training
(CDT) at the University of Cambridge funded through STFC.
This research project was conducted using
computational resources at the Maryland Advanced Research Computing Center
(MARCC). We acknowledge the Cambridge Service for Data Driven Discovery (CSD3)
system at the University of Cambridge
and Cosma7 and 8 of Durham University through STFC capital Grants
No.~ST/P002307/1 and No.~ST/R002452/1, and STFC operations Grant
No.~ST/R00689X/1. The authors acknowledge the Texas Advanced Computing Center
(TACC) at The University of Texas at Austin for providing HPC resources that have contributed to the research results
reported within this paper \cite{10.1145/3311790.3396656}. The authors would
like to acknowledge networking support by the GWverse COST Action CA16104,
``Black holes, gravitational waves and fundamental physics.’’. This research
made use of the following software: SciPy \cite{jones_scipy_2001},  Matplotlib
\cite{Hunter:2007}, NumPy \cite{van2011numpy} and yt \cite{Turk_2010}.
\bibliography{mybib}

\clearpage \appendix
\begin{figure}[t]
\begin{center}
{\includegraphics[width=1.0\columnwidth]{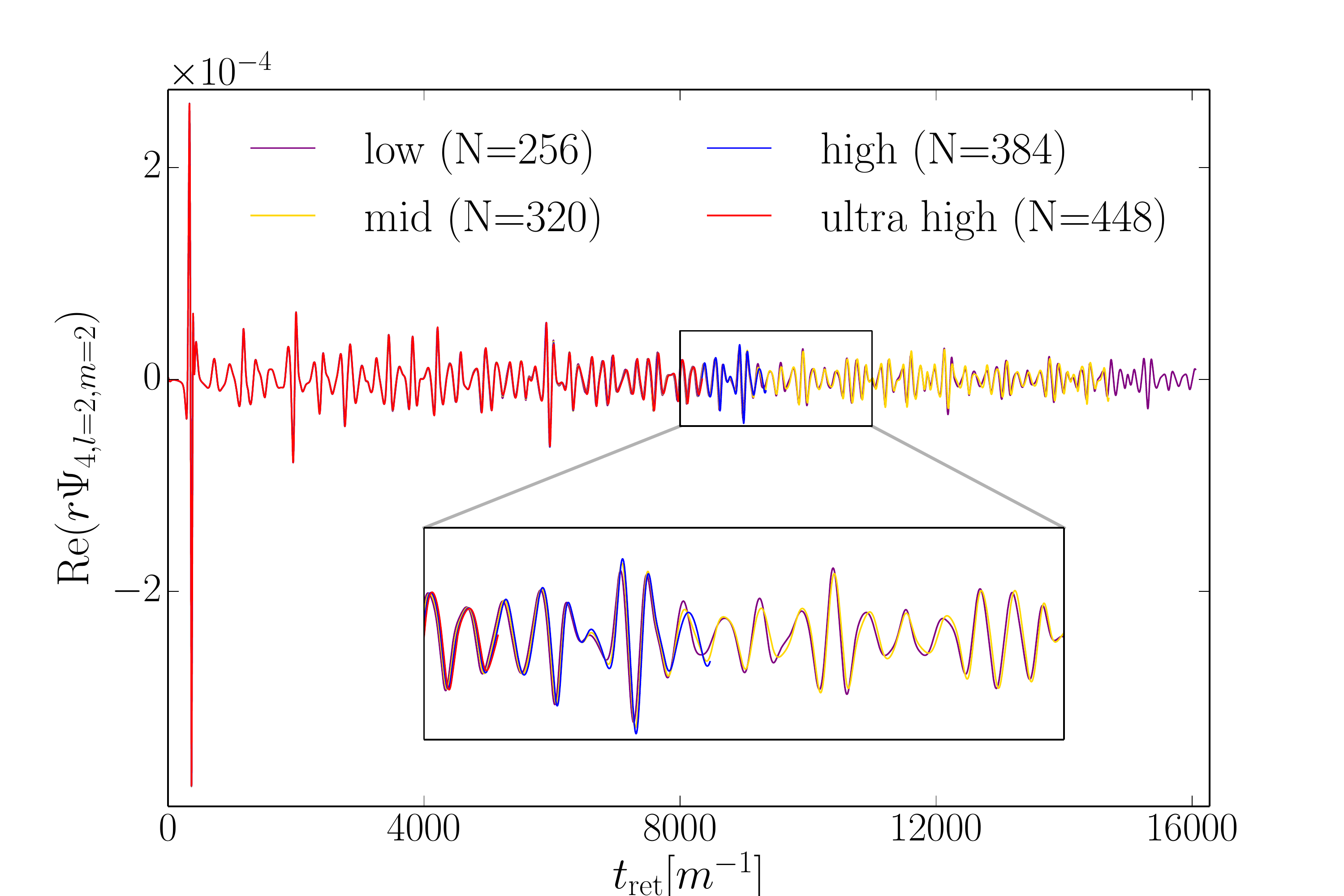}}
\caption{{\bf Convergence}: We display the $(2,2)$ multipole
of the GW signal obtained for four different resolutions
corresponding to runs 1 to 4 in \cref{tab:Overview Runs}.
A quantitative analysis yields overall convergence at
first order.
    }
\label{fig:Convergence}
\end{center}
\end{figure}
\section{Numerical Methodology}\label{appendix:numerical_methodology}

The simulations of this work have been performed with
\grchombo, a multipurpose numerical relativity code
\cite{Andrade:2021rbd,Radia:2021smk,Clough:2015sqa} which evolves the CCZ4 \cite{Gundlach:2005eh,Alic:2011gg}
formulation of the Einstein equation. The 4 dimensional spacetime metric is
decomposed into a spatial metric on a 3 dimensional spatial hypersurface,
$\gamma_{ij}$, and an extrinsic curvature $K_{ij}$, which are both evolved along
a chosen local time coordinate $t$. The line element of the decomposition is
\begin{equation}
ds^2=-\alpha^2\,{\rm d}t^2+\gamma_{ij}({\rm d}x^i + \beta^i\,{\rm d}t)({\rm d}x^j + \beta^j\,{\rm d}t)~,
\end{equation}
where $\alpha$ and $\beta^i$ denote the lapse function and shift vector. For more details about the {\sc grchombo} code and the
system of evolution equations see \cite{Radia:2021smk,Andrade:2021rbd}.

The matter part of the Lagrangian is given by
\begin{equation}
\mathcal{L}_m=-\frac{1}{2}g^{\mu\nu}\nabla_{\mu}{\bar{\varphi}}\nabla_{\nu}{\varphi}-\frac{1}{2} V(\varphi)~,
\end{equation}
whose variation with respect to the scalar field gives the evolution equation
\begin{equation}
-\nabla_\mu \nabla^\mu\varphi+\frac{\partial V(\varphi)}{\partial \vert{\varphi}\vert^2}\varphi=0~.
\end{equation}
We implement this equation as a first-order system in
terms of the CCZ4 variables in the form 
\begin{align}
&\partial_t \varphi = -\alpha \Pi +\beta^i\partial_i \varphi  ~ , \\
\begin{split} &\partial_t \Pi = \beta^i \partial_i \Pi - \chi \tilde{\gamma}^{ij}\partial_i \varphi \partial_j \alpha + \alpha\bigg( K\Pi + \frac{\partial V(\varphi)}{\partial \vert \varphi \vert^2}\varphi  \\
&\hspace{1.0cm} +  \chi\tilde{\gamma}^{ij}(\tilde{\Gamma}^k_{ij}\partial_k \varphi-\partial_i\partial_j \varphi)  + \frac{1}{2} \tilde{\gamma}^{ij}\partial_i \chi \partial_j \varphi\bigg) \end{split}\,.
\end{align}
The energy-momentum tensor is
\begin{equation}
T_{\mu\nu} =
\nabla_{(\mu}\bar{\varphi}\nabla_{\nu)}\varphi+g_{\mu\nu}\mathcal{L}_m,
\end{equation}
and its space-time projections, defined by
\begin{align} \label{eq:Mattersources}
&\rho = n_\mu \,n_\nu\,T^{\mu \nu}\,,\quad S_i = -\gamma_{i\mu }\,n_\nu\,T^{\mu \nu}\,, \nonumber \\
&S_{ij} = \gamma_{i\mu }\,\gamma_{j\nu}\,T^{\mu \nu}\,,\quad S = \gamma^{ij}\,S_{ij} ~,
\end{align}
are  
\begin{align} \label{eq:Mattereqns}
&\rho = \frac{1}{2}\Pi  \bar{\Pi} + \frac{1}{2} \partial_{i} \varphi \partial^{i} \bar{\varphi} +\frac{1}{2}V(\varphi)~, \\ 
&S_i  = \frac{1}{2}\left(\bar{\Pi} \partial_{i} \varphi + {\Pi} \partial_{i} \bar{\varphi} \right)  ~,
\label{eq:Si} \\
&S_{ij} = \partial_{(i} \varphi \partial_{j)} \bar{\varphi}  -  \frac{1}{2} \gamma_{ij}\left( \partial_{k} \varphi \partial^{k} \bar{\varphi}-\Pi  \bar{\Pi} + V(\varphi)\right) ~.
\end{align}
These are the expressions for the matter terms that source
the evolution of the spacetime geometry according to the
evolution equations (13)-(18) in Ref.~\cite{Radia:2021smk}.

\begin{figure}[ht]
\begin{center}
{\includegraphics[width=1.0\columnwidth]{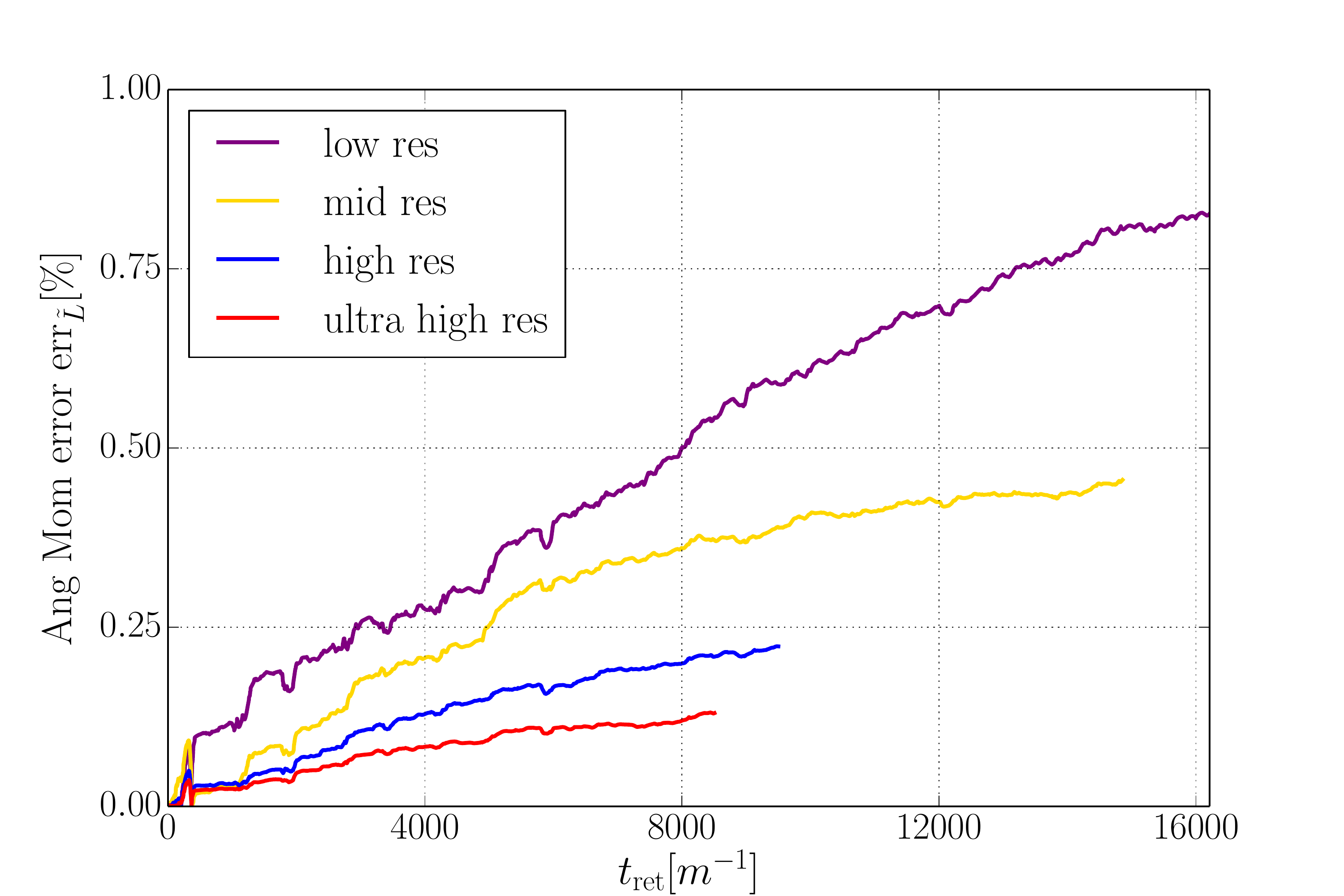}}
    \caption{{\bf Angular momentum error }: We numerically
    verify the conservation equation
    \ref{eqn:conserved_ang} for the simulations with parameters 
    given in Table \ref{tab:Overview Runs}. In the continuum limit this
    quantity becomes zero. The colored curves
    obtained for different resolutions demonstrate
    conservation with excellent accuracy of
    about $\sim 0.8\%$ error after 16000$~m^{-1}$
    and exhibit convergence to the zero continuum limit.
    }
\label{fig:ViolCons}
\end{center}
\end{figure}

\section{Angular momentum measure}\label{sec:methods}

\begin{figure*}[t]
\begin{center}
    {\includegraphics[width=1.5\columnwidth]{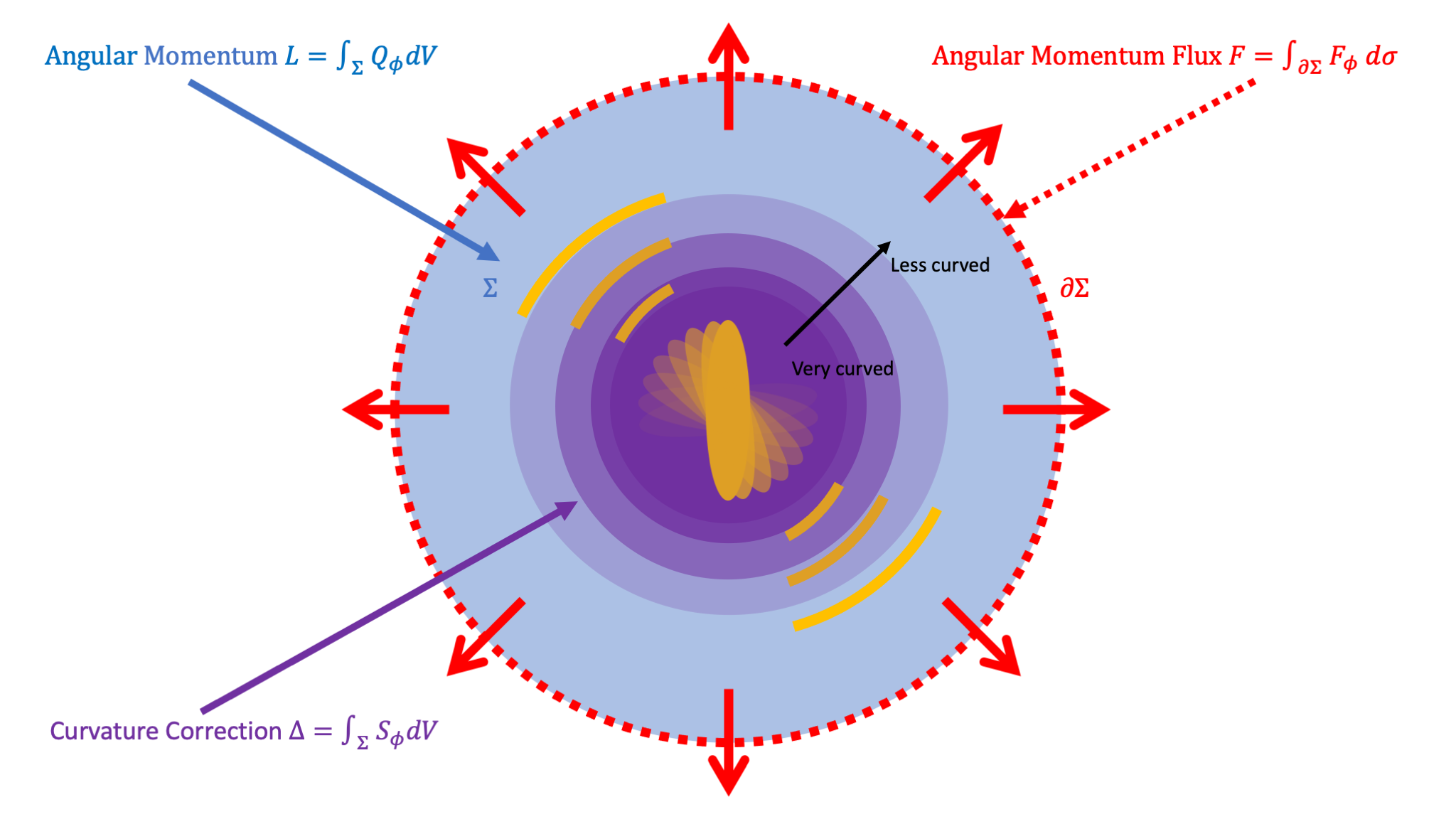}}
\caption{{\bf Schematic representation of the angular momentum measure}: We illustrate the origin of the different quantities 
appearing in the conservation equation
(\ref{eqn:conserved_ang}): The angular momentum
    ${L}$, the {angular momentum flux} $F$ and the
    curvature correction ${\Delta}$. Recall that the post-merger
    configuration does not have a toroidal structure
    as would be expected for the spinning BS model of
    Ref.~\cite{PhysRevD.90.024068,Yoshida:1997qf}, but is more bar-like in shape.
    }
\label{fig:AngMomTrick}
\end{center}
\end{figure*}

Conserved quantities in general relativity are associated with isometries of the spacetime manifold. In particular, if a spacetime conserves energy, then there must exist a time-like Killing vector field $\bs{\xi}$. A classic example is that of the Kerr vacuum solution. On the other hand, in a less symmetric spacetime like that of a black-hole merger, such a Killing vector field does not usually exist except in the asymptotically flat region. In this section we define a diagnostic quantity for the angular momentum that does not require a Killing vector, but merely a vector field to generate a measure that
converges to the classical angular momentum definition
in the flat-spacetime limit.

In order to define such a measure in a
precise manner, we will follow the work of \cite{Croft:2022gks,Clough:2021qlv} where readers will also
find more details of the calculations.

We start by defining, based on our Cartesian coordinates, the azimuthal vector
\begin{equation}
    {\xi}^\mu = (\partial_{\phi})^\mu = y(\partial_x)^\mu - x (\partial_y)^\mu~.
\end{equation}
We next define the angular angular momentum
\begin{equation}
    \mathbin{{L = \int_{\Sigma} \mathcal{Q} \sqrt{\gamma}\,
    \dd^3x}}\,,
\end{equation}
with the volume element $\sqrt{\gamma}$ on the spatial hypersurface $\Sigma$, which in our case is a sphere of finite radius, and 
\begin{align}   
    \mathcal{Q} = -S_\phi, 
\end{align}
where $S_\phi = S_i (\partial_\phi)^i$ is the azimuthal component of
the mixed space-time projection $S_i$ given by Eq.~(\ref{eq:Si}). 
We also integrate the angular momentum flux density $\mathcal{F}$ through
$\partial \Sigma$, to get the total angular momentum flux
\begin{equation}
    \mathbin{{F = \int_{\partial \Sigma} \mathcal{F} \sqrt{\sigma }\,
    \dd^2x}}~.\label{eqn:DefAngMomFlux}
\end{equation}
Here $\sqrt{\sigma}$ is the induced volume element on $\partial \Sigma$ and the flux density is
\begin{align}
    \mathcal{F} &= \frac{1}{\sqrt{\gamma^{rr}}}\left(\alpha S^r_\phi -\beta^r S_\phi\right).
\end{align}
where $\bs\beta$ is the shift vector where $S^r_\phi = {S^{\mu}}_{\nu} (\partial_r)_{\mu} (\partial_\phi)^{\nu} $ and with $\bs{\partial}_{r}$ the radial unit vector.

If $\bs{\xi} = \bs{\partial}_\phi$ is a Killing vector the rate of change of the momentum
within a given volume $\Sigma$ will be equal to the momentum flux through the
boundary, i.e.
\begin{equation} \label{eq:traditional_conservation}
    \partial_t {L} = {F}.
\end{equation}
However, in general dynamical spacetimes, we do not have such
a Killing vector and angular momentum is not conserved in this simple manner. 
Instead, we obtain a further term representing curvature contributions on the right-hand side of Eq.~(\ref{eq:traditional_conservation})
which effectively acts as a further source or sink of momentum for the matter \cite{Croft:2022gks,Clough:2021qlv}. This term is given by
\begin{equation}
    \mathbin{{\Delta = \int_{V} \mathcal{S} \sqrt{\gamma }\,
    \dd^3x}}~,
\end{equation}
where the momentum source density $\mathcal{S}=\alpha\nabla_{\mu}\xi^{\mu}$ expressed in terms
of the 3+1 variables is
\begin{align}
    \begin{split}\mathcal{S} &= \alpha S^\mu_{\nu}{}^{(3)}\partial_\mu \xi^\nu + \alpha S^\mu_{\nu} {}^{(3)}\Gamma^\nu_{\,\,\,\mu \sigma} \xi^\sigma \\ \quad &- S_\nu \beta^i \partial_i \xi^\nu  + S_\nu \xi^\mu \partial_\mu \beta^\nu - \rho \xi^\mu \partial_\mu \alpha\,, \end{split}
\end{align}
for a spacelike vector $\bs\xi$. The corresponding term
for a timelike vector $\bs\xi$ is given (see Eq.~(19) in \cite{Clough:2021qlv}).
We can thus generalize \cref{eq:traditional_conservation} to the exact conservation law
\begin{equation}
    \partial_t\mathbin{{L}} +\,
    \mathbin{{\Delta}} =
    \mathbin{F}~;\label{eqn:conserved_ang}
\end{equation} 
a schematic overview of the different contributions to this balance
law is shown in Fig.~\ref{fig:AngMomTrick}. In the flat-space limit $\gamma_{ij}
\rightarrow \delta_{ij}$, so that
find that ${{\Delta}}\rightarrow 0$ and we recover Eq.~\ref{eq:traditional_conservation} as expected given the
symmetry of flat spacetime.

In our simulations we introduce the adjusted angular momentum
defined as
\begin{equation}
 \tilde{L} = {L}+\int_{0}^t {\Delta} \,\dd t, \label{eqn:DefAngMom}~,
\end{equation}
which obeys the equation 
\begin{equation}
\label{eq:modified conservation}\partial_t\tilde{L} = {F}.
\end{equation}
We can then define a relative measure for the error in the
conservation of angular momentum as
\begin{equation}
    {\rm err }_{\tilde{L}} = \frac{\tilde{L} - \int F {\rm d}t}{{L}(t=0)}~.\label{eqn:ang_mom}
\end{equation}
The resulting error measured for our BS binary using the
four different resolutions is shown in Fig.~\ref{fig:ViolCons}
and exhibits clear convergence towards the expected limit of
zero.

We reiterate that this angular momentum measure is a
local quantity that obeys a rigorous conservation law
given by Eq.~(\ref{eq:modified conservation}) for
any chosen volume.
Its calculation does therefore not require extrapolation
to infinity (as is needed, for example, for the calculation
of the GW signal) nor even asymptotic flatness of the underlying spacetime.
However, to relate it to a more physical measure, such as the ADM angular momentum of the spacetime, we do require such conditions.
Also note that for matter fields decaying to zero on the surface of $\Sigma$ the quantity $\tilde{L}$ (but not $L$) is constant in time for a general spacetime. 

In practice, we monitor the conservation of $\tilde{L}$ as follows.
We calculate our angular momentum measure by integrating (i) the angular momentum (see \cref{eqn:DefAngMom}), and (ii) integrating the flux $F$ (see \cref{eqn:DefAngMomFlux}) over time; we set the integration constant equal to ${L}(t=0)$. Having thus obtained two measures for the same quantity allows us to estimate the uncertainty by taking the difference between the two; see \cref{fig:ViolCons} for the evolution of the error. 
Using the final value of $\tilde{L}$ obtained for Run 2
(see Table \ref{tab:Overview Runs}) inside the volume of
a coordinate sphere of coordinate radius $60~m^{-1}$, we estimate the final spin
of the merged BS as $0.0321 \pm 0.0007~\mpl^2 m^{-2}$.

\section{Numerical Accuracy}
\label{sec:numacc}

To assess the accuracy of our results we have performed simulations with four different resolution given, in terms of the number of points in each direction on the coarsest level, by $N = \{256,320,384,448\}$, all for a box width 
of $1024~m^{-1}$). As demonstrated in Fig.~\ref{fig:Convergence}, the individual wave signals obtained for this range of resolution are
in excellent agreement.
More quantitatively, we observe convergence at about 1st order convergence using the four resolutions for the $\ell = 2~m=2$ mode of $r\Psi_4$. Additionally, we estimate the discretization and error from the finite extraction region in $r\Psi_4$ and we find that the latter dominates, causing a $\sim 7\%$ error.

To calculate the energy (see \cref{fig:EnergyPower}) we used the the $\ell = 2 $ modes and using the equation for the power
\begin{equation}\label{eq:GWpower}
    \frac{{\rm d}E}{{\rm d}t} = \lim_{r\rightarrow \infty} \frac{1}{16 \pi} \sum_{\ell = 2, m} \left|\int_{t_0}^{t}  r\Psi_{4,\ell m}{\rm d}t^\prime \right|^2~,
\end{equation}
where we apply an 5th order Butterworth high-pass filter (using the scipy implementation \cite{jones_scipy_2001}) on the integral $r\Psi_{4,\ell m}$. As otherwise accumulating numerical error causes a  drift in the Energy over time. 
We estimate the error, similarly as for $r\Psi_4$, using both the discretisation error as well as the error from the finite extraction radius. For the discretisation error we use Richardson extrapolation assuming 1st order convergence to estimate the error. Similarly as before, we find that the mostly the error in the finite extraction radius dominates, giving us the value $ 0.04~\% \pm 0.0014~\% $. 

Lastly to calculate the average rate of emission, we simply perform a linear fit over the whole Energy over time signal of the largest radius of the ``medium'' run (see \cref{fig:EnergyPower}). We also perform this fit on a rolling window of $3000~m^{-1}$ to determine any reduction of radiation over time. Excluding the merger, we find the rate roughly declines by $20 \%$ over the simulation time. 

 The gravitational wave data and angular momemtum data is available at \url{https://github.com/ThomasHelfer/BosonStarAfterglow}.

\end{document}